\documentclass[letterpaper,12pt,oneside,onecolumn]{article}%
\usepackage{geometry}%
\usepackage{titlesec}%
\titleformat{\section}[hang]{\normalfont\normalsize\bfseries}{\thesection}{12pt}{\centering}%
\titleformat{\subsection}[display]{\normalfont\normalsize}{\thesubsection}{12pt}{\underline}%
\titleformat{\subsubsection}[runin]{\normalfont\normalsize}{\thesubsubsection}{12pt}{\underline}%
\newcommand{\PaperTitle}[1]{%
\begin{center}%
    \begin{large}%
        \textbf {#1} \\%
    \end{large}%
\end{center}%
}%
\newcommand{\AuthorList}[1]{%
\begin{center}%
    {#1} \\%
\end{center}%
}%
\newcommand{\AuthorAffiliations}[1]{%
\begin{center}%
    {#1} \\%
\end{center}%
}%
\newcommand{\Keywords}[1]{%
\begin{center}%
   Keywords: {#1} \\%
\end{center}%
}%

\usepackage[numbers,sort&compress]{natbib}
\usepackage{graphicx}

\usepackage{textcomp}
\usepackage{amsmath}
\usepackage{amssymb}
\usepackage{bm}
\usepackage{mathrsfs}
\usepackage{color}
\usepackage{hyperref}

\usepackage{array,color}

\begin{document}%
\PaperTitle{Replica-exchange Wang--Landau sampling: Pushing the limits\\
  of Monte Carlo simulations in materials sciences}%
%
%
\AuthorList{Dilina Perera$^1$, Ying Wai Li$^2$, Markus Eisenbach$^2$, 
Thomas Vogel$^3$, and David P. Landau$^1$}%
\AuthorAffiliations{$^1$Center for Simulational Physics,
The University of Georgia, Athens, GA 30602, USA\\
$^2$National Center for Computational Sciences,
Oak Ridge National Laboratory,\\ Oak Ridge, TN 37831, USA{\renewcommand{\thefootnote}{\fnsymbol{footnote}}\footnote{Notice: This manuscript has been authored by UT-Battelle, LLC, under Contract No.
DE-AC05-00OR22725 with the U.S. Department of Energy. The United States Government
retains and the publisher, by accepting the article for publication, acknowledges
that the United States Government retains a non-exclusive, paid-up, irrevocable,
world-wide license to publish or reproduce the published form of this manuscript, or
allow others to do so, for United States Government purposes.}%
\setcounter{footnote}{0}
}\\
$^3$Theoretical Division (T-1), Los Alamos National Laboratory,
Los Alamos, NM 87545, USA}%


\section{Abstract} 

We describe the study of thermodynamics of materials using
replica-exchange Wang--Landau (REWL) sampling, a generic framework for
massively parallel implementations of the Wang--Landau Monte Carlo
method. To evaluate the performance and scalability of the method, we
investigate the magnetic phase transition in body-centered cubic (bcc)
iron using the classical Heisenberg model parametrized with first
principles calculations. We demonstrate that our framework leads to a
significant speedup without compromising the accuracy and precision
and facilitates the study of much larger systems than is possible with
its serial counterpart.


\section{Introduction}

Computational methods have become indispensable for enhancing
our understanding of material properties and making accurate predictions
of their behavior.  This can, in turn, lead to the development of improved
materials with, e.g., increased strength and resistance to
degradation.  Even with 
recent advancements, first-principles based approaches to the
electronic structure are restricted to the calculation of ground state
properties of small systems and are incapable of predicting the
system behavior at finite temperature. Instead, classical
Monte Carlo (MC) simulations based on atomistic or coarse grained
Hamiltonians have become a viable and powerful approach for probing
the thermodynamic properties of large-scale, complex
systems~\cite{landauBook}. The Metropolis method~\cite{metropolis} is
the basis of many modern MC schemes where one probabilistically
samples from the canonical (NVT) ensemble by generating a sequence of
microstates according to Boltzmann weights.  An inherent downside of
Metropolis sampling, just as for NVT molecular dynamics simulations,
is its tendency to get trapped in local free-energy minima that are
frequently encountered in systems with complex free energy landscapes. To
circumvent this pitfall, different methodologies have been
introduced. One popular approach is to perform multiple Metropolis
simulations at different temperatures and allow periodic
conformational (replica) exchanges between them.  This is widely known
as parallel tempering or replica-exchange Monte
Carlo~\cite{partemp1,partemp3}. Another approach
strives to sample microstates from a flat probability distribution (the
so-called multicanonical ensemble~\cite{muca1,muca2}), where
states have non-Boltzmann weights that are inversely proportional to the
\textit{a priori} unknown density of states. Such
generalized-ensemble methods hope to
escape free energy barriers and avoid long time scales which are encountered at low temperatures and near phase transitions.
To this end, the Wang--Landau (WL) method~\cite{wl1,wl2,wl3} has
emerged as a simple but powerful technique. The underlying idea behind
Wang--Landau sampling is to perform a random walk in configuration
space while iteratively adjusting the density of states $g(E)$ and
hence the simulation weights. This ultimately leads the accumulation
of a uniform histogram in energy $(E)$ space (or, for that matter, in
any other reaction coordinate or collective variable). Eventually, the
WL scheme delivers an estimator for the density of
states so that  
the thermodynamic behavior of the system for the
entire temperature range of interest can be extracted from a single simulation. The method has been successfully
applied to a wide array of intricate problems in condensed matter and
statistical physics including spin glasses, liquid crystals, polymers,
protein folding etc.~\cite{spinGlass, liquidCrystal, polymer,
proteinFolding}.

Multiple attempts have been made to improve the efficiency of the
Wang--Landau method, either by finding more efficient
trial moves, or by accelerating its convergence
(see~\cite{zhou05pre,belardinelli07pre} for examples). Nevertheless, in
order to exploit the power of modern high performance computing
systems and adapt the method to even larger-scale systems with
increased complexity, an efficient parallelization scheme is
essential. Recently, a generic framework was proposed, combining
replica-exchange with Wang--Landau sampling into a massively parallel
Monte Carlo simulation method \cite{rewl1, rewl2, rewl3, rewl4}. The
strategy is to split the energy range into a series of smaller,
overlapping windows, which are sampled by independent random walkers
with occasional replica-exchanges. Parallelism is then achieved mainly
in a ``divide-and-conquer'' manner.

In this paper, we review this recently introduced
parallelization scheme for WL simulations.  To demonstrate its
applicability to problems in material science, we investigate the
ferromagnetic-paramagnetic phase transition in body centered cubic
(bcc) iron via a classical spin model frequently used for describing
magnetic materials, namely the Heisenberg ferromagnet.  Furthermore,
we will elaborate on the specific advantages of the parallel method
and assess its scalability for large-scale simulations involving
hundreds of thousands of spins, utilizing thousands of
processors. This work serves as a precursor to the study of more
realistic, and more complex, model Hamiltonians for bcc iron, where
lattice vibrations contribute extra degrees of freedom and complicate
the phase space along with the energy landscape~\cite{yin12}. The
complexity of the problem restricts the study to small
system sizes using the serial WL method or traditional parallel
methods such as parallel tempering.


\section{Model}

In the Heisenberg model, atomic magnetic moments are represented as
classical spin vectors of unit length.  In the absence of magnetic
anisotropies or external fields, the total energy for a system of
interacting classical spins is given by the Hamiltonian
\begin{equation}
  \mathcal{H} =  - \sum_{i<j} J_{ij} \mathbf{S}_i \cdot \mathbf{S}_j\,,\quad
  |\mathbf{S}_i|=1\;\;\forall i\,,
\end{equation}
where $\mathbf{S}_i$ is the orientation of the $i${th} spin, and
$J_{ij}$ is the exchange interaction between $i${th} and $j${th} spin.
For the realistic modeling of magnetic properties of bcc iron, we
choose numerical values of $J_{ij}$ from the first principles
based parameterization given in Ref.~\cite{Ma2008} which limits the
interactions to the nearest and next nearest neighbors.  The possible
energies of the system lie within the range $[-N\epsilon_0,
N\epsilon_0]$, where $\epsilon_0 \approx 0.14$\,eV is the absolute
value of the ground state energy per spin, and $N = 2L^3$ is the total
number of spins. There are two spins in each unit cell and $L$ is the
lattice size, i.e., the number of unit cells in each direction.  As
our primary interest is the thermodynamic behavior near the critical
temperature for the ferromagnetic--paramagnetic transition, we impose
minimum and maximum energy cutoffs and sample configurations within a
restricted global energy range, $[E_{\text{min}}, E_{\text{max}}]$.  This
reduction drastically reduces the computational cost of sampling rare
configurations at extremely low and high energies, which potentially
becomes a bottleneck in serial WL simulations. The reduced energy
range is still large enough though that the random walkers are able to
visit all contributing microstates; therefore, this reduction
does not introduce any systematic errors.


\section{The Replica-Exchange framework for Wang--Landau sampling}

\subsection{Background in statistical mechanics}

The density of states in energy, $g(E)$, measures the energy degeneracy of
admissible states of a system, from which the partition function $Z$
can be calculated:

\begin{equation}
Z(T) = \sum_{\bm{X}}{e^{-E[\bm{X}] / k_BT}} = \sum_{E}{g(E)\, e^{-E /k_BT}},
\label{partitionFunction}
\end{equation}

\noindent
where $\bm{X}$ stands for a state or configuration that the system can
reside in; $k_B$ is the Boltzmann constant and $T$ is the temperature.
The first sum runs over all possible states of the system, whereas
the second sum runs over all possible total energies and is calculable
once $g(E)$ is known. While $g(E)$ is temperature independent
and only depends on the definition of the Hamiltonian, Eq.\
(\ref{partitionFunction}) allows for the calculation of the
temperature dependent $Z$ via the corresponding Boltzmann factors.  An
important consequence is the possibility of calculating the
thermodynamic quantities at \textit{any} temperature with the sole
knowledge of $g(E)$. For example, the average energy $\langle E
\rangle$ and the heat capacity $C_V$ can be calculated:
\begin{align}
\langle E \rangle(T) = \frac{1}{Z}\sum_{E}{g(E)\,E\, e^{-E / k_BT}}\,,\qquad
C_V(T) = \frac{\langle E^2\rangle - \langle E \rangle^2}{k_BT^2}\,.
\end{align}
The specific heat is defined as $C_V / N$. These thermodynamic
observables provide a measure to identify and locate phase transitions
and hence understand critical phenomena. For the Heisenberg
ferromagnetic iron model we use in this study, we successfully capture
the Curie temperature which corresponds to the
ferromagnetic-paramagnetic transition.

\subsection{The original Wang--Landau algorithm}

The original Wang--Landau sampling \cite{wl1,wl2,wl3} is classified as
one of the generalized-ensemble MC methods where the final outcome of
a simulation is an estimation of $g(E)$. Depending on the parameter
space in which the random walk is performed, the density of states
could as well be an estimate of the degeneracy of states as a function
of other order parameters or reaction coordinates or even a
higher-dimensional combination of such, see~\cite{2d_wl} for an example.

At the beginning of each Wang--Landau simulation the desired total
energy range, $E \in [E_{\textrm{min}}, E_{\textrm{max}}]$, for which
$g(E)$ should be obtained is determined or estimated. For a model with
continuous energy domain an energy bin width $\delta E$ is also chosen
to control the resolution of $g(E)$. An initial guess of $\tilde{g}(E)
= 1$ for all energies is used as a starting point, although other
choices are suitable as well.\footnote{Here, $\tilde{g}(E)$ denotes
  the instantaneous estimator for $g(E)$ which changes during the
  course of the simulation.} A histogram, $H(E)$, is introduced to
keep track of the number of visits to each energy; all entries are
set to zero at the beginning. The simulation starts with an arbitrary
initial configuration, and new states $B$ are generated by applying a
Monte Carlo trial move (see below) to the current configuration $A$
with energy $E_A$. The energy $E_B$ of the trial state is then
evaluated and the new state will be accepted according to the
acceptance probability:
\begin{equation}
  P(A \to B) = \min{\left[ 1, \frac{\tilde{g}(E_A)}{\tilde{g}(E_B)} \right]}.
\end{equation}
If trial state $B$ is accepted, $\tilde{g}(E)$ and $H(E)$ will be
updated by $\tilde{g}(E_B) \to \tilde{g}(E_B) \times f$ and $H(E_B)
\to H(E_B) + 1$. Otherwise, the update is done for 
the current state $A$. Here, $f$ is the modification factor whose 
initial value is $f_0 = e^1$. Trial states are continuously generated and
$\tilde{g}(E)$ and $H(E)$ are updated, until a ``flat'' histogram is
achieved, i.e. all $H(E)$ entries are no less than a certain
fraction $p$ of its average value $\bar{H}(E)$ [$H(E) \geq p
\bar{H}(E) \; \forall E$]. At this point, the modification factor is
reduced via $f \to \sqrt{f}$, the histogram is reset [$H(E) = 0$] but
$\tilde{g}(E)$ is carried over to the next iteration. This procedure
is repeated until $f$ reaches a predefined, terminal value
$f_{\textrm{min}}$.  Hence, $\tilde{g}(E)$ is modified by smaller and
smaller amounts and eventually converges (asymptotically) to the true
density of states $g(E)$.\footnote{Convergence
  proofs can be found in Ref.~\cite{zhou05pre} and, for a directly
  related method~\cite{junghans14jctc}, in Ref.~\cite{dama14prl}.}

\subsection{The parallel scheme}

The recently introduced parallel implementation of Wang--Landau
sampling~\cite{rewl1, rewl2} incorporates the original, serial
Wang--Landau algorithm into a replica-exchange framework. The global
energy range $[E_{\textrm{min}}, E_{\textrm{max}}]$ is broken into a
number ($h$) of smaller windows, each of which overlaps with its
nearest neighbors on both sides with an overlapping ratio $o$
(a~schematic diagram is shown in the upper panel of
Fig.~\ref{fig:DOS_and_Cv}a).  Multiple ($m$) Wang--Landau walkers can
be employed in each of the windows to reduce statistical errors
during the simulation~\cite{rewl2}, but each walker has its own
$\tilde{g}(E)$ and $H(E)$ and all of them must proceed to the
next iteration independently. Replica exchanges are proposed between
walkers $i$ and $j$ from adjacent windows after a 
predefined number of MC sweeps (MCS) and are accepted according to the probability:
\begin{equation}
P_{\textrm{RE}} = \min{\left[ 1, \frac{\tilde{g_i}(E_A) \tilde{g_j}(E_B)}{\tilde{g_i}(E_B) \tilde{g_j}(E_A)} \right]},
\end{equation}
where we have assumed that $A$ (or $B$) is the current configuration
of walker $i$ (or $j$).

The simulation produces multiple, overlapping fragments of $g(E)$.
They are joined at points where the slopes of $\ln
g(E)$ (i.e., $d \ln g(E) / dE$, the inverse microcanonical
temperature) best coincide. This practice reduces the introduction of
artificial kinks in the combined $g(E)$ due to the joining process
and eliminates resultant errors in thermodynamic quantities. See
Refs.~\cite{rewl3,rewl4} for illustrations.

\subsection{Simulation details}

To compare results for the parallel and serial WL schemes, we chose a
relatively small system size ($L=10$) and a global energy range
($-260$\,eV$\leq E\leq 120$\,eV) that are accessible by the serial
method within a fairly reasonable time.  For checking the convergence
of $g(E)$, we used the parameter values from the original WL paper, an
$80 \%$ flatness criterion and final modification factor of
$\text{ln}\,f_{\text{min}} = 1 \times 10^{-8}$.  For the trial updates
we used the simplest MC trial move, i.e., we randomly choose a spin and
assign a arbitrary new direction.

We decided that eleven windows ($h=11$) and replica exchanges between
neighboring windows proposed every $500$ MCS (one MCS is equivalent to
$N$ MC trial moves) are reasonable choices for the lattice size
$L=10$. Since replica exchanges can, by construction, only be accepted
if both corresponding replica are in the energy overlap region between
two windows, small overlaps lead to low acceptance rates whereas
unnecessarily large overlap downgrades the performance.  We found that
a moderately large overlap of $o \approx 75 \%$ can be a good choice
for maintaining a balance between fast convergence and reasonable
acceptance rates~\cite{rewl2}.  With this choice of the overlap, we
observed acceptance rates in the range of $52-55\%$ for replica
exchanges.


\section{Results and discussion: Accuracy and scalability}

\begin{figure}[b!]
\centering
  \includegraphics[width=0.87\textwidth]{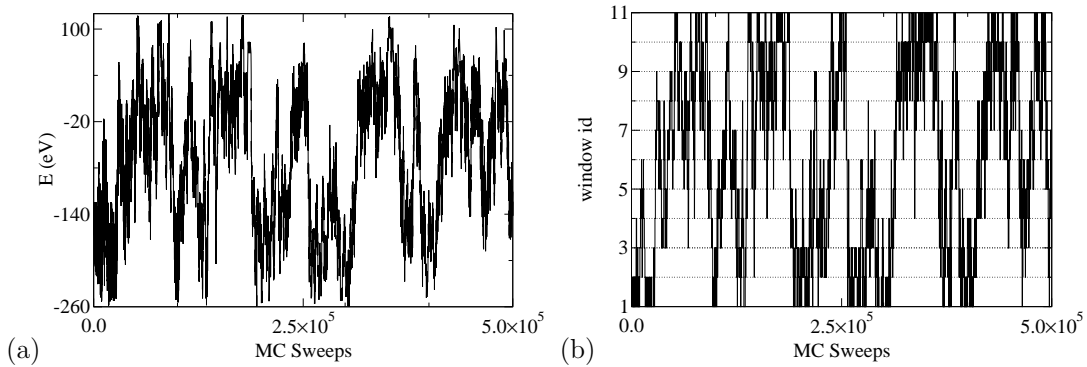}   
  \caption{Path of a single replica through the entire energy space
    (a) and energy windows~(b) during the first $5 \times 10^5$ MCS.
    Replica exchanges are attempted every $500$ MCS, acceptance rates
    are in the range of $52-55\%$.  The replica completes a round-trip
    approximately every $7 \times 10^4$ MCS.  }
 \label{fig:path_of_rep}
\end{figure}

With those chosen values of $h\!=\!11$ and $o\!=\!75\%$, we performed parallel
simulations for $L\!=\!10$ while employing a single walker per window
($m\!=\!1$) for simplicity. Fig.~\ref{fig:path_of_rep} shows the time
series of a single replica as it performs smooth round-trips across
the entire energy range (a) and through all the energy windows (b).
Similar behavior was observed for all replicas, verifying 
that the random walkers are not restricted to certain regions of the
energy space.

\begin{figure}[t]
  \includegraphics{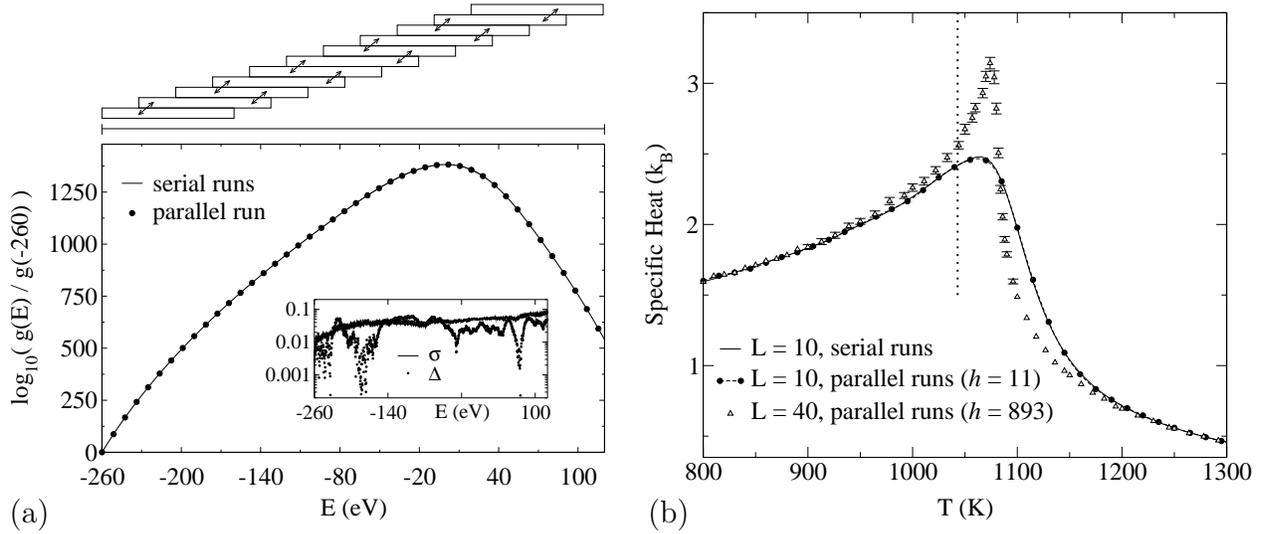}
  \caption{ (a) Logarithm of the density of states for the lattice
    size $L=10$.  The solid line represents the average of $10$
    independent serial runs, filled dots show results from a single
    parallel run.  The inset illustrates
    the accuracy of the parallel method.  Solid line in the inset
    shows the standard deviation $\sigma$ of the serial results while
    the dots show the absolute numerical difference $\Delta$ between
    the serial and parallel results.  (b)~Specific heat curves as
    functions of temperature for $L=10$ and $L=40$.  The dotted line
    marks the experimental Curie temperature, $T_\mathrm{C} =
    1043$\,K.  For $L=10$, the results from the serial runs are shown
    for comparison.}\vskip-.5\baselineskip
 \label{fig:DOS_and_Cv}
\end{figure}

Fig.~\ref{fig:DOS_and_Cv} (a) compares the density of states obtained
from both parallel and serial runs.  Data shown by filled dots were
obtained from a single parallel simulation, the solid line represents
the average of $10$ independent serial runs, the standard deviation
$\sigma$ is shown in the inset.  The absolute difference $\Delta$
between the results from the serial runs and the parallel run is also
shown in the inset for comparison.  $\Delta$ is of the same size as
$\sigma$.  Thus, the precision of the parallel WL results are on par
with those from conventional, serial runs.
To accommodate the increase in the global energy range for much larger
systems, we increased the number of windows accordingly while keeping
the energy window size ($\Delta E \approx 108.57$\,eV) and the overlap
($o=75\%$) constant. For example, for $L=40$ ($-16640$\,eV $\leq E\leq
7680$\,eV), $893$~windows were required.  While a serial WL simulation
would take years to estimate the density of states, the parallel
scheme achieved this within two days.  Fig.~\ref{fig:DOS_and_Cv} (b)
shows the specific heat derived from the estimated density of states
for the system sizes $L=10$ and $L=40$. Despite the simplicity of our
model, the peak positions match reasonably well with the experimental
Curie temperature $T_\mathrm{C} = 1043$\,K~\cite{TcExp}.  For comparison, we
also show the results obtained from the serial WL simulations for
$L=10$.  The data from the serial and parallel runs are within the
mutual error bars.

\begin{figure}[b!]
  \includegraphics{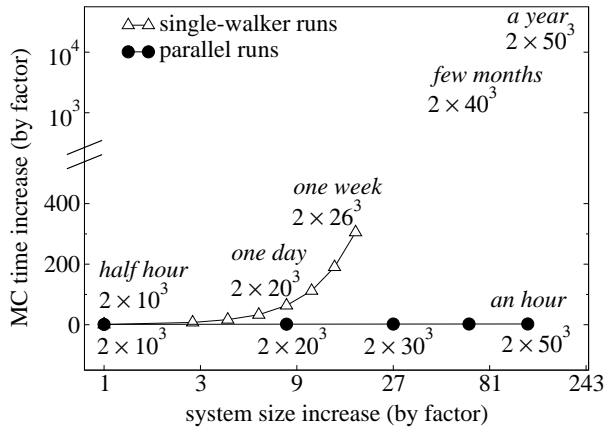}\hfill
  \begin{minipage}[b]{17pc}
    \caption{Relative increase in simulation time versus the system
      size increase.  Filled dots are for parallel runs while the open
      triangles represent serial runs.  Simulation time is determined
      by the time needed to complete the first WL iteration.  The
      number of energy windows $h$ were increased with the increasing
      system size while keeping the size of the windows and the
      overlap fixed.}
    \label{fig:weak_scaling}
  \end{minipage}
\end{figure}
The success of any parallel scheme depends not only on its accuracy
and precision, but also on the scalability, i.e., the ability to
effectively utilize an increasing number of processors.  Ultimately,
by increasing the number of computing cores one hopes to achieve two
types of scaling behavior: to speed up the execution while fixing the
problem size (strong scaling), or to increase the problem size without
increasing execution time (weak scaling).  Here, we will demonstrate
the weak scaling behavior of our method using the Heisenberg model as
a test case.  For this analysis, we performed REWL simulations for
lattice sizes $10\leq L\leq50$, (corresponding to spin numbers
$2000\leq N\leq 250000$ respectively), and measured the number of MC
trial moves needed to complete the first Wang--Landau iteration.  The
results of this study are summarized in Fig.~\ref{fig:weak_scaling},
where we compare the simulation time for serial runs (triangles) with
the parallel performance (dots).  For the serial runs, simulation time
increases dramatically with system size whereas for the parallel runs
it remains almost constant.


\section{Summary}

The successful application of replica-exchange Wang--Landau (REWL)
sampling to a Heisenberg bcc iron model in the study of magnetic phase
transition proves its immense potential in obtaining finite
temperature properties of materials via classical Monte Carlo
simulations. The excellent agreement between REWL and the original,
serial WL results, along with the nearly perfect weak scaling with the
number of cores used, means that heretofore inaccessible large scale
studies are now possible. With a model correctly parametrized by first
principles calculations (using a relatively small system size), we can
bridge different length scales between first principles calculations
(typically in nanoscale) and model simulations (mesoscale or beyond)
in computational physics/materials sciences.  This will eventually
enable the study of finite temperature properties of materials having
the relevant length scales needed for real world applications.

\paragraph{Acknowledgments}

{{We wish to thank T. W\"ust for stimulating discussions.
    This research was sponsored in part by the U.S. Department of
    Energy, Office of Basic Energy Sciences, Materials Sciences and
    Engineering Division (M.E.), by the ``Center for Defect Physics'',
    an Energy Frontier Research Center (D.P.), and by the Office of
    Advanced Scientific Computing Research (Y.W.L.). This research
    used resources of the Oak Ridge Leadership Computing Facility at
    ORNL, which is supported by the Office of Science of the U.S.
    Department of Energy under Contract No. DE-AC05-00OR22725.
    LA-UR-14-27450 assigned.}}


\end{document}